# A Novel Graph-based Computation Offloading Strategy for Workflow Applications in Mobile Edge Computing

Xuejun Li, *Member, IEEE,* Tianxiang Chen, Dong Yuan, *Member, IEEE,* Jia Xu, *Student Member, IEEE,* and Xiao Liu, *Senior Member, IEEE*

**Abstract**—With the fast development of mobile edge computing (MEC), there is an increasing demand for running complex applications on the edge. These complex applications can be represented as workflows where task dependencies are explicitly specified. To achieve better Quality of Service (QoS), for instance, faster response time and lower energy consumption, computation offloading is widely used in the MEC environment. However, many existing computation offloading strategies only focus on independent computation tasks but overlook the task dependencies. Meanwhile, most of these strategies are based on search algorithms such as particle swarm optimization (PSO), genetic algorithm (GA) which are often time-consuming and hence not suitable for many delay-sensitive complex applications in MEC. Therefore, a highly efficient graph-based strategy was proposed in our recent work but it can only deal with simple workflow applications with linear (namely sequential) structure. For solving these problems, a novel graph-based strategy is proposed for workflow applications in MEC. Specifically, this strategy can deal with complex workflow applications with nonlinear (viz. parallel, selective and iterative) structures. Meanwhile, the offloading decision plan with the lowest energy consumption of the end-device under the deadline constraint can be found by using the graph-based partition technique. We have comprehensively evaluated our strategy using both a real-world case study on a MEC based UAV (Unmanned Aerial Vehicle) delivery system and extensive simulation experiments on the FogWorkflowSim platform for MEC based workflow applications. The evaluation results successfully demonstrate the effectiveness of our proposed strategy and its overall better performance than other representative strategies.

---------- ◆ ----------

# 1 INTRODUCTION

WITH the continuous improvement of the computing capacity of various smart end-devices, an increasing number of intelligent applications are deployed on mobile end-devices such as smart traffic, smart healthcare, smart logistics and so on. Meanwhile, massive computation requests submitted by the end-devices can be handled by cloud datacentres with unlimited computing resources. However, massive data transmission over public networks with limited bandwidth will cause significant delays, which is unacceptable for many delay-sensitive applications. Nowadays, mobile edge computing (MEC) has been widely used to provision computing resources from the network edge to the end-device in order to reduce response delay [1]. Computation tasks on the end-devices can be offloaded to the edge servers for execution through low-cost and high-bandwidth transmission such as the 5G and WIFI networks [2]. Computation offloading plays a key role in effectively improving the QoS of MEC-based applications by reducing the response delay and the energy consumption of end-devices [3], [4].

Given the success of MEC, there is an increasing demand for running complex applications on the edge. For example, in the UAV (Unmanned Aerial Vehicle) based smart delivery system, there are many complex applications such as dynamic route planning, obstacle detection and face recognition [5]. These applications are important parts of the whole delivery process. However, like most smart end-devices, UAVs are limited by their computing power and battery life so that they are unable to execute computation-intensive tasks as mentioned above. Fortunately, the UAV's energy consumption and task response time can be effectively reduced by the computation offloading technology in the MEC environment [6]. Meanwhile, as will be shown in the motivating example in the next section, most complex applications in the real-world can be rep-resented by workflows where task dependencies are explicitly specified [7]. However, currently many studies only focus on independent tasks without considering task dependencies. A few of them considered simple linear task dependencies where tasks are executed in a sequential manner [8], [9], [10]. Generally speaking, all real-world applications can be represented by a mix of linear (namely sequential) and nonlinear (viz. parallel, selective, and iterative) structures [11]. Therefore, computation offloading for complex applications should be

- X. Li, T. Chen and J. Xu are with the School of Computer Science and Technology, Anhui University, Hefei, Anhui, China. E-mail: xjli@ahu.edu.cn; biyisi_96@qq.com; xujia@stu.ahu.edu.cn.
- D. Yuan is with the School of Electrical and Information Engineering, University of Sydney, Sydney, NSW 2006, Australia. E-mail: dong.yuan@sydney.edu.au.
- X. Liu is with the School of Information Technology, Deakin University, Geelong, Australia. E-mail: xiao.liu@deakin.edu.au.

(Corresponding author: Xiao Liu.)

able to deal with both linear and nonlinear structures. Meanwhile, given the complex nature of the computation offloading problem, the greedy type of strategies have been widely used to obtain a feasible solution in a short time [12], but they cannot produce optimal offloading decisions. Therefore, in order to improve the quality of the decisions, many studies employed search algorithms such as particle swarm optimization (PSO) and genetic algorithm (GA) to search for the optimal offloading decision by iterative process, which could produce significant time overhead [13], [14], [15], [16]. In [17], Hu et al., proposed a learning-driven algorithm to achieve efficient offloading decision plans, which need lots of prior data training to get an effective model. Therefore, at present, most computation offload-ing strategies are either simple but not good enough, or they are too time-consuming to be suitable for delay-sensitive complex applications [18],[19].

To address the above issues, a novel computation offloading strategy using graph partition technology is proposed in this paper for workflow applications in MEC. To distinguish with our previous preliminary work which can only deal with linear work-flow structures [20], we name the previous strategy Graph4Edge-Linear, name the new strategy proposed in this paper Graph4Edge-Nonlinear. Our proposed strategy considers the influence of the complex task dependencies on the computation offloading decisions, and the end device's energy consumption is optimized effectively under the given deadline constraints. Please note that the energy consumptions of edge servers are not considered in this paper. This is because edge servers are usually connected to the power grid, and hence their energy consumptions are not regarded as limiting factors in a MEC environment.

Specifically, the novel contributions of this paper are summarized as follows:

1) A novel nonlinear workflow model for complex MEC-based applications is proposed. The model is based on WDG (Workflow Dependency Graph) which considers both complex task dependencies and the objective of reducing the end-device's energy consumption.

2) We propose a novel graph-based computation offloading strategy named Graph4Edge-Nonlinear based on the WDG which can find the best computation offloading decision with the minimum end-device' energy consumption under the given deadline. Its performance is significantly better than popular search-algorithm based strategies.

3) Both a case study on a real-world UAV delivery system [21] and extensive simulation experiments on the FogWorkflowSim platform for MEC based workflow applications [22] are presented. The experimental results demonstrate the effectiveness of our proposed strategy and its overall better performance than other representative strategies.

The rest of this paper is structured as follows: Section 2 introduces a motivating example on a MEC-based UAV delivery system. Section 3 presents some preliminaries for this study. Section 4 proposes our novel graph-based com-putation offloading strategy for workflow applications with nonlinear structures. Section 5 presents the evaluation results. Section 6 reviews the related works for computations offloading. Finally, Section 7 makes the conclusions and points out some future work.

## 2 MOTIVATION EXAMPLE AND PROBLEM ANALYSIS

An example of the MEC-based UAV last-mile delivery scenario is presented to describe the problem of computation offloading in the MEC environment in this section.

In the MEC-based UAV last-mile delivery system, there are various delay-sensitive applications such as dynamic flight route planning and autonomous obstacle avoidance for UAVs, pose and face recognition for receivers [23], [24]. These applications usually consist of object detection, pattern recognition and video stream processing, which are computation intensive tasks. In fact, because of the UAV's limited battery life and computing power, these computation intensive tasks are not suitable for executed locally under the fast response and energy efficiency requirements. Therefore, computation offloading to the edge server is often required.

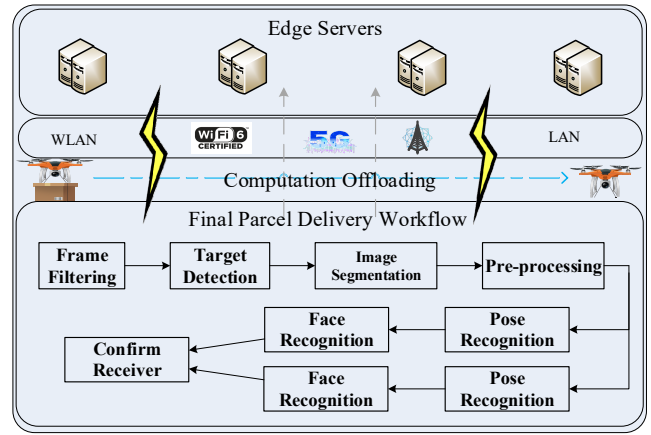

Fig. 1. Computation offloading for an example workflow in a MEC-based UAV delivery system

Here, we illustrate the computation offloading problem with a partial workflow of the whole UAV delivery process, namely the final parcel delivery workflow. The MEC-based UAV delivery system is conceptually divided into two layers. As shown in Fig. 1, the upper layer is the edge server layer, which consists of various edge servers. These edge servers can provide computing resources close to the UAV. The bottom layer is the final parcel delivery workflow which consists of a set of computation tasks with dependencies. Generally speaking, dependencies can be generated in two situations. The first situation is where a task has one or multiple predecessors and/or successor tasks, namely they have temporal dependencies. The second situation is where data transfer is required between the two neighbouring tasks, namely they have data dependencies. In this paper, we simply refer them as task dependencies, and they can be explicitly specified using DAG (Directed Acyclic Graph) as will be introduced in the next section. Specifically, as shown in Fig. 1, there are many computation-intensive tasks in the UAV last-mile delivery scenario, for example, target detection, image segmentation, pose and face recognition. Specifically, in the real world, the video frames may contain multiple persons and hence it is necessary to segment the images and run multiple poses and face recognition tasks in parallel to ensure timely detection of the actual receiver from the crowd. Once the UAV detects the actual receiver, it will approach the receiver and begin to land and unload the parcel. Obviously, these real-time tasks are delay-sensitive, and fast response time is essential.

According to different characteristics (such as task workload, data size and deadline constraints) of the computation tasks, some of them are offloaded to edge servers to achieve better QoS such as faster response time and lower energy consumption [25]. However, computation offloading is a difficult decision-making problem.

While there are some existing strategies which are based on heuristic algorithms or search algorithms, they all have some limitations. For example, heuristic algorithms have the premature convergence issue so that they may not be able to find the best computation offloading decision. While search algorithms such as particle swarm optimization (PSO) and genetic algorithm (GA) can find the best decisions in theory, they are usually very time-consuming and hence not suitable for delay-sensitive applications. Meanwhile, to the best of our knowledge, none of the existing strategies can effectively deal with complex task dependencies which can be represented by nonlinear workflow structures such as parallel, selective and iterative structures.

For solving the above issues, a novel graph-based strategy is proposed to solve the computation offloading problem in the MEC environment. Our proposed strategy can deal with nonlinear workflow structures and find the best computation offloading decision with the minimum energy consumption under the deadline constraint.

## 3 PRELIMINARIES

Generally speaking, for the purpose of computation offloading, there are two kinds of computation tasks in the workflow, which are general tasks and local execution tasks. General tasks are those tasks which are executed either at end-device or edge server via computation offloading. Local execution tasks are those tasks that can only be executed on end-device due to the required input data is only available at the end-device and cannot be moved due to security restrictions, or some tasks which require user input at the end-device [20]. In this situation, edge servers cannot handle these tasks. In other words, these tasks must be processed on the end-device.

This paper uses the workflow dependency graph (WDG) to represent the workflow model and its task dependencies. WDG is a directed acyclic graph (DAG) that is composed of workflow tasks with dependencies. Each task $T_i$ of WDG contains three basic attributes $x_i, y_i, z_i$, which represent the energy consumption in different situations of the end-device.

In Fig. 2, the symbol $\rightarrow$ denotes that there is a dependency relationship between two task nodes. For example, the $T_i \rightarrow T_j$, indicates that $T_i$ is the predecessor of $T_j$ in the WDG. There are $T_1 \rightarrow T_2$, $T_2 \rightarrow T_3$, $T_2 \rightarrow T_5$, $T_3 \rightarrow T_4$, $T_5 \rightarrow T_6$, etc. $T_1$ points to $T_2$, which means there is a direct dependency between $T_1$ and $T_2$. We use the task $T_1$'s output data as task $T_2$'s input data. In addition, $\rightarrow$ is defined as having transitivity, where $T_i \rightarrow T_k \rightarrow T_j \Leftrightarrow T_i \rightarrow T_k \wedge T_k \rightarrow T_j \Leftrightarrow T_i \rightarrow T_j$.

The symbol $\leftrightarrow$ indicates that there is no dependency between the two tasks, where $T_i \leftrightarrow T_j$ means the $T_i$ and $T_j$ are disparate branches in WDG. For instance, we have $T_3 \leftrightarrow T_5$, $T_4 \leftrightarrow T_6$, etc. in Fig. 2.

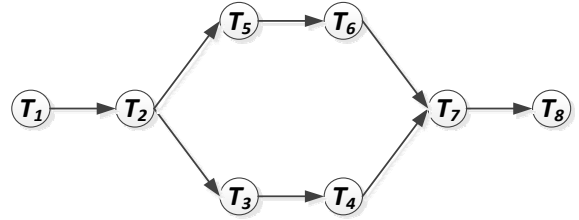

Fig. 2. An example workflow dependency graph

Here, $x_i$ means the energy consumption of the data transmission when decide offloading $T_i$ to edge server. The data transmission between $T_i$ and $T_j$ (MB) indicated as $Comm(T_i, T_j)$. $T_i$'s direct predecessor node is $T_j$. $Bandwidth$ is data transfer speed of the computation tasks (Mbps). $P_{trans}$ denotes the end-device's transmission power (W).

$$x_i = \frac{Comm(T_i, T_j)}{Bandwidth} * P_{trans} \tag{1}$$

Where $y_i$ is end-device's idle energy consumption when decide offloading $T_i$ to edge server. The task $T_i$'s workload (Megacycles) is denoted as $l_i$. The edge server's CPU frequency (GHz) is $f_{edge}$. The end-device's idle power (W) is $P_{idle}$.

$$y_i = \frac{l_i}{f_{edge}} * P_{idle} \tag{2}$$

The end-device's load energy consumption is denoted as $z_i$ when $T_i$ is executed on the end-device. The end-device's CPU frequency (GHz) and execution power (W) is denoted as $f_{end}$ and $P_{end}$ respectively.

$$z_i = \frac{l_i}{f_{end}} * P_{end} \tag{3}$$

In our previous work [20], the proposed method can convert the computation offloading problem of the linear WDG into the shortest path problem. There are two types of virtual nodes in the WDG, which are the start node and end node in problem of the shortest path. The weight of the edge between two nodes is expressed as $w < T_i, T_j >$, which is the end-device's energy consumption. At this time, the computation task $T_i$ and $T_j$ are decided offloading to edge server and the tasks between $T_i$ and $T_j$ are executed locally. All possible offloading decisions in WDG can be mapped to edges between different task nodes. The weight of each edge represents the energy consumption of end-devices. As mentioned before, the energy consumptions of edge servers are not considered as edge servers are usually connected to the power grid.

According to the WDG, an Energy consumption Transitive Graph (ETG) is constructed. Specifically, we design a one-to-one mapping for all paths in the graph to all possible offloading decisions of the workflow, and the classic Dijkstra algorithm is used to find the shortest path in ETG. Since the complexity of graph-based algorithm is low, this strategy can efficiently obtain excellent results.

# 4 GRAPH-BASED MINIMUM ENERGY CONSUMPTION COMPUTATION OFFLOADING STRATEGY

We present the strategy of Graph4Edge-Nonlinear in this section. This strategy will find the best offloading decision with minimum energy consumption of the end-device under the given deadline. First in Section 4.1, we introduce our model definition and how to convert workflow structure to WDG. Then in Section 4.2, we discuss how to find the best offloading decision plan for complex WDG with Graph4Edge-Nonlinear. Finally, in Section 4.3, we use pseudo-code to describe the detailed process for our strategy and the discussion of its algorithm complexity is also presented.

## 4.1 Problem Formulation

We introduce the model definition in this section. Then the conversion from nonlinear workflow structure to WDG is described.

### 4.1.1 Model Definition

For the computation offloading purpose, the attributes of task $T_i$ are defined as $\langle x_i, y_i, z_i, flag_i, E_i \rangle$. $flag_i$ denotes the constraint whether $T_i$ is local execution task or not. Specifically, $flag_i = 1$ means that $T_i$ is a local execution task. Otherwise, the task $T_i$ is sending to the edge server for execution. $E_i$ denotes the $T_i$'s execution energy consumption in end-device. The calculation method of $E_i$ as follows:

$$E_i = \begin{cases} x_i + y_i & , \text{if } T_i \text{ is offloaded and } T_{i-1} \text{ is not;} \\ x_i + z_i & , \text{if } T_{i-1} \text{ is offloaded and } T_i \text{ is not;} \\ y_i & , \text{if } T_i \text{ and } T_{i-1} \text{ are offloaded;} \\ z_i & , \text{if } T_i \text{ and } T_{i-1} \text{ are not offloaded;} \end{cases} \quad (4)$$

According to the offloading decision of $T_i$ and $T_{i-1}$, the calculation method of $E_i$ has two situations. When either of them offloaded, the data transmission appears between the two tasks. In this situation, the end device's energy consumption depends on $T_i$'s task offloading decision plan. When both of them are offloaded or not offloaded, there is no data transmission between the two tasks. The end device's energy consumption depends on offloading decision plans of two tasks. According to the above definition, the optimization goal is defined as $MinEC$, which is calculated as follows:

$$MinEC = \min \sum_{\{T_i | T_j \in WDG\}} (E_i) \quad (5)$$

The final optimization goal is to find the best computation offloading plan with the minimum energy consumption of the end-device under the deadline constraints.

As shown in Fig. 3, the WDG has a sub-branch within one block. If the task offloading decision with minimum energy consumption in WDG can be found, only one branch is chosen to construct the ETG which is called "main branch" (indicated as MB). The rest of branches are called "sub-branches" (indicated as SB). such as the $MB = \{T_1, T_2, T_3, T_4, T_7, T_8\}$ and $SB = \{T_5, T_6\}$ in Fig. 3. The energy consumption of the SB in the block is mapped to the weight of the MB. The weight of the SB in Fig. 3 is defined as $E_5 + E_6$. To better present our problems and methods, here are some detailed definitions.

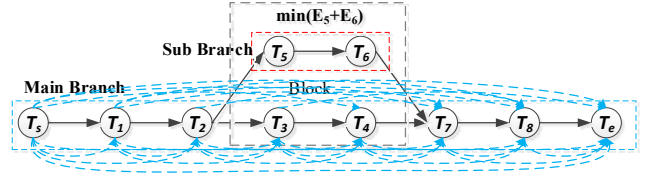

Fig. 3. An example of building ETG for single-block WDG

*Block*: In WDG, A block (denoted as B) is defined as a set of sub-branches which are forked from one task node and merged at another task node. A WDG with simple block $B = \{T_3, T_4, T_5, T_6\}$ is shown in Fig. 3.

*In-block edge:* In-block edge $e\langle T_i, T_j \rangle$ represents the edge begins with $T_i$ preceding the block, and points to $T_j$ in the block, for example, $e\langle T_1, T_3 \rangle$, $e\langle T_1, T_4 \rangle$ in Fig. 3. Formally, $e\langle T_i, T_j \rangle$ is an in-block edge, where $\exists T_k \in WDG \wedge T_i \to T_k \wedge T_k \to T_k$.

*Out-block edge:* Out-block edge $e\langle T_i, T_j \rangle$ represents the edge begins with $T_i$ in the block, and points to $T_j$ succeeding the block, for example, $e\langle T_3, T_8 \rangle$, $e\langle T_4, T_8 \rangle$ in Fig. 3. Formally, $e\langle T_i, T_j \rangle$ is an out-block edge, where $\exists T_k \in WDG \wedge T_i \to T_k \wedge T_j \to T_k$.

*Over-block edge:* Over-block edge $e\langle T_i, T_j \rangle$ represents the edge crosses over the block. $T_i$ is the task node which precedes the block and $T_j$ is the task node which succeeds the block. For example, $e\langle T_1, T_8 \rangle$, $e\langle T_2, T_8 \rangle$ in Fig. 3. Formally, $e\langle T_i, T_j \rangle$ is an over-block edge, where $\exists T_k, T_h \in WDG \wedge T_i \to T_k \to T_j \wedge T_i \to T_k \wedge T_h \leftrightarrow T_k$.

*Ordinary edge:* An ordinary edge $e\langle T_i, T_j \rangle$ means that tasks between $T_i$ and $T_j$ when they are totally ordered, such as $e\langle T_1, T_2 \rangle$, $e\langle T_3, T_4 \rangle$, $e\langle T_7, T_8 \rangle$ in Fig. 3. Formally, $e\langle T_i, T_j \rangle$ is an ordinary edge, where $\neg \exists T_k \in WDG \wedge \left( (T_i \to T_k \wedge T_k \leftrightarrow T_j) \vee (T_i \leftrightarrow T_k \wedge T_k \to T_j) \vee (T_h \in WDG \wedge T_h \leftrightarrow T_k \wedge T_i \to T_h \to T_j \wedge T_i \to T_k \to T_j) \right)$.

In this paper, we can define the weight of the ordinary edge as Eq. (6).

$$w(T_i, T_j) = y_j + x_{i+1} + x_j + \sum_{\{T_k | T_k \in WDG \wedge T_i \to T_k \to T_j\}} z_k \quad (6)$$

### 4.1.2 Convert Nonlinear Workflow to Complex WDG

The topology of workflow is represented by WDG. In the real-world, the structure of workflow contains four basic topology types, viz. sequential, parallel, selective and it-erative structures [11]. Specifically, in this paper, we name the workflow only composed with the sequential structure as the linear workflow. If it contains the other three structures, we name the workflow as the nonlinear structure. Although the real-world workflow structures can be very complex with the mix of the four basic workflow structures, all of them can be converted to WDGs by a simple method are proposed in [26]. In [20], a strategy that can find minimum energy consumption of a simple linear structure workflow is proposed. However, the task dependencies in the remaining three workflow structures are much more complicated. Through the conversion process, any structures can be converted into multiple sequential structures. Fig. 4 shows how the three nonlinear basic workflow structures are converted to WDG.

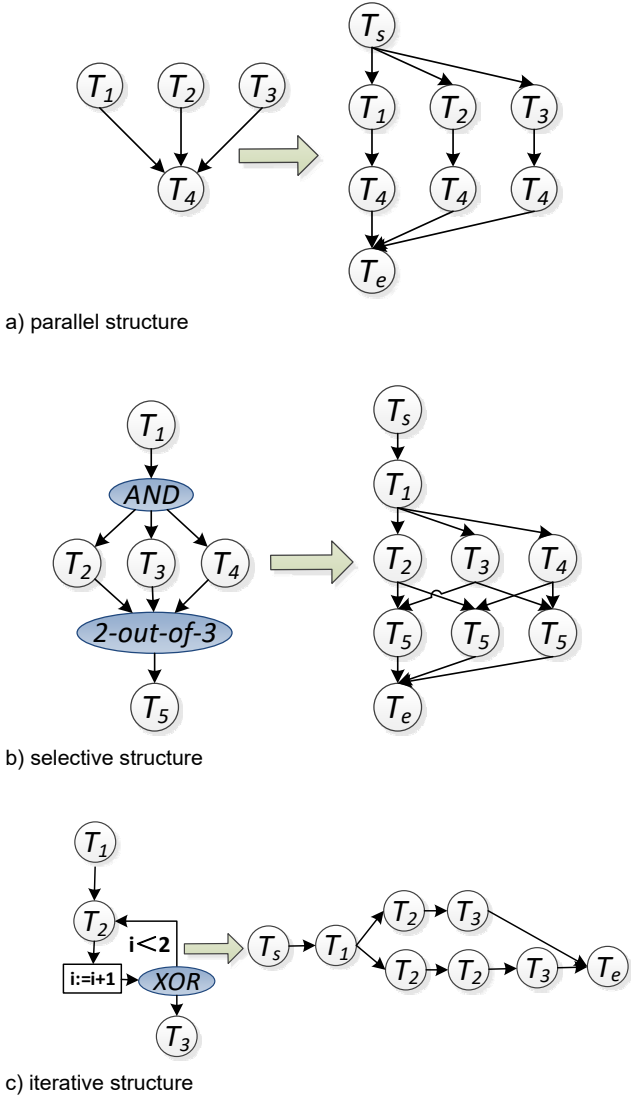

a) parallel structure

b) selective structure

c) iterative structure

Fig. 4. Convert nonlinear workflow structures to WDG

As shown in Fig. 4 (a), three sequential structures are obtained by constructing three subtask instances for a parallel structure [27]. The virtual start node and end node are added to connect these sequential structures to form a complete WDG. Fig. 4 (b) and (c) show the conver-sion examples for the selective and iterative structures, respectively [7].

Any workflow is a combination of the four basic structures, and they can be converted into corresponding WDG models. In real-world workflow applications, WDGs with nonlinear structures are very common [28]. Due to the existence of nonlinear structures, the existing computation offloading strategy for linear structures cannot be used directly. For solving the problem of computation offloading for complex workflow applications with nonlinear structures, we propose the Graph4Edge-Nonlinear strategy. This strategy is able to optimize the end-device's energy consumption under the given deadline constraints.

### 4.2 Graph4Edge-Nonlinear for Complex WDG

#### 4.2.1 Single-block WDG

In this section, the single-block WDG is analyzed as an example to describe the problem and the detailed steps for Graph4Edge-Nonlinear based on the above model definition.

The purpose of the Graph4Edge-Nonlinear strategy is to map the energy consumption of the possible offloading decision to the weight of the edge. Therefore, the shortest path of the linear workflow structure can be easily found by the optimal offloading strategy. $e\langle T_i, T_j \rangle$ means that tasks $T_i$ and $T_j$ are offloaded to the edge server, and tasks between $T_i$ and $T_j$ are executed locally. As a result, it is necessary to calculate the energy consumption of locally executed tasks which include transmission and idle energy consumption. In the single-block WDG, Eq. (6) is suitable for in-block edges and ordinary edges. But when $e\langle T_i, T_j \rangle$ is either out-block or over-block, Eq. (6) is no longer suitable for its weight calculation due to the tasks succeeding the block may have more than one task as their predecessor node. For example, the edge of $\langle T_3, T_8 \rangle$ in Fig. 3 can be calculated by $w\langle T_3, T_8 \rangle = x_4 + x_8 + y_8 + z_4 + z_7$ according to Eq. (6). However, the offloaded situation of the sub-branch tasks $T_5$ and $T_6$ is not considered. Therefore, the obtained shortest path cannot represent the overall decision of WDG.

For the above reasons, the weight of $e\langle T_i, T_j \rangle$ is defined as follows:

$$w(T_i, T_j) = y_j + x_j + x_{i+1}^*$$
$$+ \sum_{\{T_k \mid T_k \in WDG \wedge T_i \to T_k \to T_j\}} z_k + \left( \sum_{\{T_l \mid T_l \in SB\}} E_{T_l} \right)_{S_{min}} \quad (7)$$

In Eq. (7), $\left( \sum_{\{T_l \mid T_l \in SB\}} E_{T_l} \right)_{S_{min}}$ means the minimum energy consumption of the tasks that are in the sub-branches of the block. $x_{i+1}^*$ represents the transmission energy consumption of the task that directly depends on $T_i$. The out- or over-block edge's shortest path length is equal to the task's minimum energy consumption by Eq. 7, i.e. $P_{min}\langle T_i, T_j \rangle = \sum_{\{T_k \mid T_k \in WDG \wedge T_i \to T_k \to T_j\}} E_k$. Hence, in order to calculate the out- or over-block edge's weights, the offloaded strategy of sub-branch in single-block WDG is essential. For instance, the weight of edge $e\langle T_3, T_8 \rangle$ in Fig. 3 is calculated as $w\langle T_3, T_8 \rangle = x_4 + x_8 + y_8 + z_4 + z_7 +$ $(E_5 + E_6)_{S_{min}}$ where $S_{min}$ is the best computation offloading decision of the SB.

For any sub-branch, the offloading decision of the previous node of the block decides the transmission energy consumption of the first task node in SB. If $e\langle T_i, T_j\rangle$ is an over-block edge, we need to consider two different situations. If $T_i$ is the previous node of the block and it will be offloaded, then when the first task of the SB is offloaded, the transmission energy consumption is not calculated as they are both executed on the edge. Otherwise, the transmission energy consumption must be calculated when the first task of the SB is offloaded. In order to solve this problem, when $T_i$ is not the previous node of the block, a special non-offloaded virtual node $T_s'$ is added between the start node $T_s$ and the first task node $T_1$, and set $x_s = y_s = z_s = 0$. In this way, we account the energy consumption for transmission of the first node of the SB. If the rest of the tasks in the block compose a linear WDG, the Graph4Edge-Linear strategy can find its minimum energy consumption decision. Otherwise, if the remaining tasks within the block still compose a complex WDG with nonlinear structures, the Graph4Edge-Nonlinear strategy must be recursively called to search the best offloading decision plan for sub-branches.

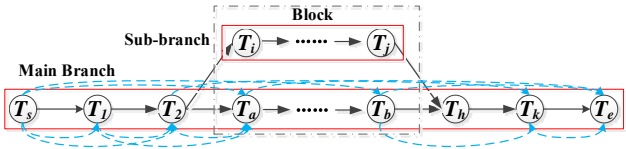

Fig. 5. The initial ETG of single-block WDG

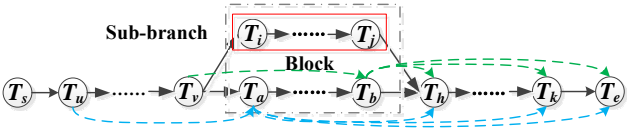

Fig. 6. Examples of in-block edges in two different situations

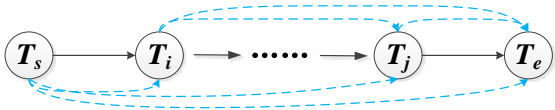

(a) In-block edges not from the previous node

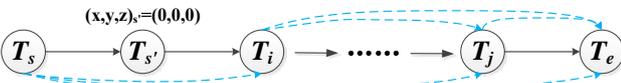

(b) In-block edges from the previous node

Fig. 7. Construct the ETG for branch structure

For an out-block edge $e\langle T_i, T_j\rangle$, the previous task of sub-branch is unknown. For example, for calculating the out-block edge weight $e\langle T_3, T_8\rangle$ in Fig. 3, the optimal offloading decision $S_{min}$ is necessary for the sub-branch $\{T_5, T_6\}$. However, $S_{min}$ depends on the offloaded status of $T_1$ and $T_2$. Therefore, multiple ETGs for WDG must be constructed to measure the out-block edge's weight. Spe-cifically, the minimum energy consumption strategy is the minimum length path among all ETGs. The specific steps for Graph4Edge-Nonlinear strategy are shown below:

**Step 1**: Construct the initial ETG of WDG. An arbitrary branch in WDG is chosen as the main branch. At the same time, the energy edges are added to construct ETG. And for the set of $\{T_i | T_i \in localSet\}$, the edges are pruned when $T_i$ serves as the head or tail.

**Step 2**: Set the weight of edges in the ETG. The weights of the ordinary and in-block edges are set by Eq. (6). For the over-block edges, the Graph4Edge-Nonlinear strategy is recursively called to find its $S_{min}$, then set the weights by Eq. (7). Finally, the weight of all out-block edges is set to infinity. The initial ETG is shown in Fig. 5.

**Step 3**: Construct two different branch ETG models based on in-block edge situations. The specific description is as follows:

1) *If the in-block edge is not from the previous node of the block, and firstly discovered.* A new ETG is created, and then the Graph4Edge-Nonlinear strategy processes the sub-branch in the block to find the optimal energy consumption offloading decision. For example, when we find $e\langle T_u, T_a\rangle$ in Fig. 6, mark the current situation and create a new ETG to record according to the current ETG. First of all, the information of the current ETG is copied to the new ETG. Then, we prune all the in-block edges which head from the previous task of this block, which ensures the correct calculation of the sub-branch's minimum energy consumption strategy. At this time, the ETG generated by the linear branch WDG is shown in Fig. 8 (a). Finally, the weights of all out-block edges for this block in ETGs are updated.

2) *If the in-block edge is from the adjacent predecessor task of block, and this situation was first discovered.* We can make adjustments in the current ETG. Specifically, when the in-block edges from the previous task node of this block were discovered for the first time, such as $e\langle T_v, T_b\rangle$ in Fig. 6, situation (1) has been completely traversed, so it can be processed directly on the current ETG. Prune all in-block edges that are not from the previous task node, and it ensures that all sub-branches can be directly calculated using the Graph4Edge-Nonlinear strategy. At this time, the ETG created by the linear branch WDG is shown in Fig. 8 (b). Finally, all out-block edge weights are updated for this block in ETGs.

**Step 4**: Use the Dijkstra algorithm to search the minimum length path in ETGs, and perform verification to ensure that deadline constraints are met. The nodes on the shortest path are the minimum energy consumption offloading strategy have found.

### 4.2.2 Multiple-blocks WDG

In real workflow-based applications, WDG's structures can be complex with multiple blocks in the WDG. Therefore, Graph4Edge-Nonlinear strategy should be able to deal with multiple-blocks in the WDG.

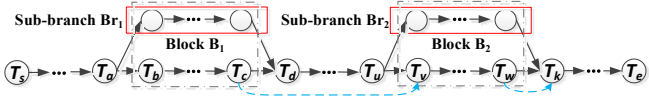

Fig. 8. WDG with multiple serial blocks

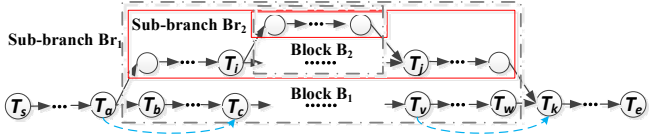

Fig. 9. WDG with nested branches

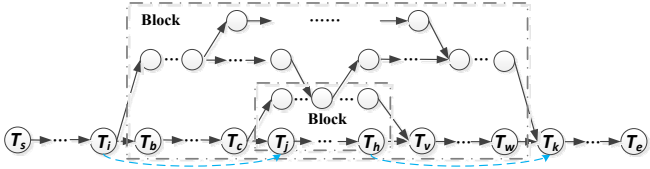

Fig. 10. The ETG of WDG

A WDG may consist of many blocks. At first, any branch can be selected as the main branch. This main branch is used to construct the initial ETG. Then, multiple ETGs are built for different blocks. In the calculation process of out-block and over-block edge weights, two new situations need to search minimum energy consumption offloading decision for the sub-branch.

1) *WDG with multiple serial blocks.* In this situation, there is an edge that is both an out-block edge for one block and an in-block edge for another block, e.g. $e\langle T_c, T_v \rangle$ in Fig. 8. In our strategy, depending on the head node of the in-block edge for B1, the offloaded strategy with the sub-branch is different. As a result, both the head node and the in-block edge weight for $Br_2$ will change. In order to calculate the out-block edge weight for B2, e.g. $e\langle T_w, T_k \rangle$, the offloading strategy of $Br_2$ for B2 must make sure, which depends on the offloading strategy of SB1 for B1. So it is necessary to find its minimum energy consumption offloading strategy from $Br_1$ of B1.

2) *WDG with nested branches.* In this situation, it is necessary to recursively call the Graph4Edge-Nonlinear strategy to find its optimal offloading strategy. For example, $e\langle T_a, T_c \rangle$ in Fig. 9 is an in-block edge of blocks B1 and B2, multiple new ETGs should be created based on the different situations of the two blocks, to find the optimal offloading strategy of sub-branches $Br_1$ and $Br_2$. Hence it is necessary to recursively call the Graph4Edge-Nonlinear strategy for the WDG $Br_1 \cup Br_2$.

The ETG for an example complex WDG is shown in Fig. 10. By recursively calling the Graph4Edge-Nonlinear strategy for the sub-branches, the minimum energy consumption offloading decision of the whole WDG can be found. For example, given an in-block edge $e\langle T_i, T_j \rangle$ in Fig. 10, the Graph4Edge-Nonlinear strategy calculates the sub-branch $\{T_u | T_u \in WDG \land T_u \to T_k \land T_u \leftrightarrow T_j \land T_u \leftrightarrow T_h\}$, and gets the weight of out-block $e\langle T_h, T_k \rangle$.

### 4.2.3 Strategy description

Clearly, no matter how complicated the structure of the WDG is, it can always be transformed to the linear structure by calling Graph4Edge-Nonlinear strategy recursively. Here, we present the pseudo-code for the Graph4Edge-Nonlinear strategy.

---

**Strategy**: Graph4Edge-Nonlinear

**Input**: A workflow dependency graph ($WDG$);
　　　local-execution tasks in WDG ($localSet$);
　　　The workflow task's deadline constraint ($Deadline$);

**Output**: tasks in WDG ($S$);

1　Compute $(x_i, y_i, z_i)$ for all tasks by Eq. (1-3);
2　Add $T_s, T_e$ into WDG and set attributes;
3　**if** $T_s^{'}$ is needed **then**
4　　Add a virtual task $T_s^{'}$ succeed $T_s$ and set attributes;
5　**end if**
6　**if** $WDG$ is linear workflow **then**
7　　return Graph4Edge-Linear ($WDG$, $localSet$, $Deadline$);
8　**end if**
9　Get a main branch $MB$ from $WDG$ and construct ETG;
10　Prune $e(T_i, T_j)$ if $T_i$ or $T_j$ in $localSet$;
11　Set all out-block edges $e(T_i, T_j) = \infty$;
12　Compute weight for other edges by Eq. (6-7);
13　$ETG\_Set = ETG_{init}$;
14　**for** each in-block edge $e(T_i, T_j)$ in $ETG\_Set$ **do**
15　　**if** $isFirstFind(T_i)$ && $notPreviousNode(T_i)$ **then**
16　　　$ETG_{Set} \leftarrow ETG$;
17　　　$S_{SB} \leftarrow Iterate$ Graph4Edge-Nonlinear;
18　　**end if**
19　　**if** $isPreviousNode(T_i)$ **then**
20　　　$S_{SB} \leftarrow Iterate$ Graph4Edge-Nonlinear (need $T_s^{'}$);
21　　**end if**
22　　Compute out-block edge in $ETG\_Set$ by Eq. (7);
23　**end for**
24　**for** $k = n + 1$ to 1 **do**
25　　$P_{min} = Dijkstra\_Algorithm(T_s, T_k, ETG)$;
26　　$S = P_{min}(T_s, T_k)$ traversed tasks;
27　　**if** $workflow$ $makespan < Deadline$ **then**
28　　　break;
29　　**end if**
30　**end for**
31　**if** $k = 0$ **then**
32　　$S = null$;
33　**end if**
34　return $S$;

---

First, WDG is initialized (Lines 1-5). If the WDG is a linear structure, the Graph4Edge-Linear strategy is directly called (Line 7). Otherwise, an arbitrary branch from $T_s$ to $T_e$ is chosen as the main branch, and this main branch is used to construct the initial ETG (Lines 9-10), and compute the weight of the ordinary, in-block and over-block edge by Eq. (6-7) (Line 12). Next, all the in-block edges are traversed in sequence. When the in-block edge of a block is found for the first time, and its head node is not the previous task for this block, a new ETG is created and added to the $ETG\_Set$ (Lines 15-18). Then, the shortest path of the sub-branch can be found in the new ETG. When an in-block edge of which the head node is the previous task of this block is found, the current ETG is processed (Lines 19-21) to obtain the sub-branch's optimal offloaded strategy. Meanwhile, the weight of the out-block edges is updated in $ETG\_Set$(Line 22). Finally, the Dijkstra algorithm can find the shortest path from $T_s$ to $T_k$ (Line 25). In the worst situation (Line 31), if all solutions do not meet the deadline constraints, all tasks in the workflow are executed locally.

For the pseudo-code in Graph4Edge-Nonlinear, recursive calls (Lines 14-23) exist in the Graph4Edge-Nonlinear strategy, and the complexity of the algorithm highly depends on the structure of WDG. The worst time complexity of initial ETG (Lines 1-5) is $O(n^2)$. For each branch, a new ETG will be created for WDG (Line 16). The created ETG's maximum number of tasks is smaller than the number of tasks in the main branch, which is n. For all ETGs, the construction of directed edges needs to be completed, and the worst time complexity of this operation is $O(n^3)$. Use the Dijkstra algorithm to perform the shortest path search for all ETGs when the deadline constraint is met (Line 25), its time complexity is $O(n * n^2 * n)$. In summary, the worst time complexity is $O(n^4)$. Similarly, for each ETG, multiple two-dimensional arrays store the necessary information, so that the space complexity of the Graph4Edge-Nonlinear strategy is about $O(n^3)$, and does not exceed $O(n^4)$.

# 5 EVALUATION

The Graph4Edge-Nonlinear strategy is able to find the optimal computation offloading decision with the lowest energy consumption under the given deadline constraint for a complex workflow application. In this section, we describe the simulation environment and parameter settings firstly. Then, we revisit our motivating example as a real-world case study to illustrate our strategy's effectiveness. Furthermore, with simulation experiments based on a real-world UAV delivery system UAV-EXPRESS, we evaluate the performance of Graph4Edge-Nonlinear and compare with other strategies in the end-device's energy consumption, the strategy running time and task response time. UAV-EXPRESS is developed based on EX-PRESS[1] which is an energy-efficient and secure framework for MEC environment and blockchain technology-based smart systems [21].

## 5.1 Case study

Similar to the motivating example shown in Fig.1, Fig. 11 shows the detailed workflow for the final parcel delivery process in the UAV delivery system. Before the UAV reaches the destination for parcel delivery, the edge server downloads the facial images of the parcel receiver from the cloud server of the logistics system. When the UAV arrives at the destination, the video stream captured by the camera of the UAV is analyzed frame by frame to locate the position and confirm the identity of the parcel receiver.

The final parcel delivery process can be described in three stages. In the first stage (including Tasks 1-3), the original video frames are filtered on the UAV. We use the target detection function to search for video frames containing people, and frames without people are filtered directly. A small proportion of the images containing people in the whole video frame is further extracted by using the image segmentation algorithm according to the detected location coordinate of the people. In the second stage (including Task 4 and Task 5), extracted video frames are further processed. In a real-world scenario, there could be a lot of pedestrians near the destination, and hence there will be many images containing multiple people. We use pose recognition (i.e., the parcel recipient will receive an instruction on her/his mobile App to make a specific pose such as waving the right hand from right to left) to further identify the receiver from the crowd. Finally, in the third stage (including Task 6 and Task 7), identification of the parcel receiver is conducted using face recognition. Face recognition confirms whether the person waving hand matches the face images downloaded in advance from the cloud server. If matches, the UAV will approach the receiver for landing and then hand over the goods. Otherwise, more image frames are processed until the correct parcel receiver is found, or the delivery process terminates without successfully locating the parcel receiver.

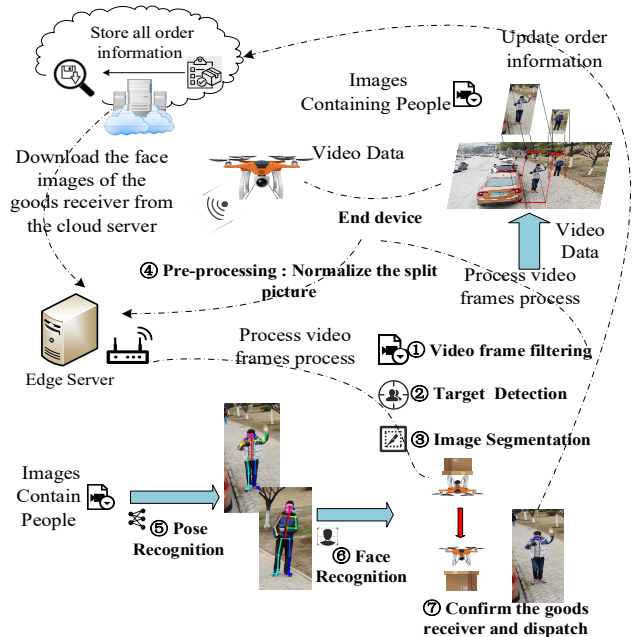

Fig. 11. A case study on the MEC-Based UAV delivery system

For the experiment, we select a one-minute video clip, with 1920*1080 pixel and 60 FPS recorded by the DJI Mavic Air UAV[2]. With our computation offloading strategy, the generated computation offloading decision is that Task 1-3 and Task 4 are to be executed locally, while Task 5 and Task 6 are to be offloaded to the edge server. Specifically, Task 1-3 and Task 4 are the video pre-processing tasks that can be executed on the UAV since the characteristic of data size but the required computing power is low. After the UAV completes the video pre-processing tasks, the remaining data size is reduced to 10% of the original video frames. The deep neural network-based computation tasks (Task 5 and Task 6) are executed at the edge server because the size of data transfer for Task 5-6 is very small but they both require high computing power.

## 5.2 Simulation Environment and Parameter Settings

In our experiments, we compare Graph4Edge-Nonlinear with three types of computation offloading strategies. The first type is based on search algorithms including PSO and GA which are most widely used for computation offloading. The second type is a Greedy strategy which makes the offloading decision for each task through comparison the energy consumption required for offloading with local execution. If the energy consumption of task offloading is less than execution locally, then the task will be offloaded to the edge server for execution. Otherwise, the task will be executed locally. The third type is the All-in-End strategy, which means that all of the tasks are executed in the end device. All computation offloading strategies are applied with the Min-Min task scheduling algorithm [29] at the edge server.

All simulation experiments are implemented on the FogWorkflowSim platform, which is a simulation platform for Fog/MEC-based workflow applications [22]. It supports different kinds of workflow structures and different evaluation index metrics such as time, energy and cost. The experiments are run on a laptop with the following configuration: Intel® Core™ I7-9750H CPU 2.60GHz, 16G RAM, NVIDIA GeForce GTX 1660Ti.

### TABLE 1
### MEC ENVIRONMENT PARAMETER SETTING

| Parameters | End Server | Edge Server |
|---|---|---|
| MIPS | 1000 | 1300 |
| Load Power (mW) | 700 | N/A |
| Idle Power (mW) | 30 | N/A |
| Data Transmission Power (mW) | 100 | N/A |

Table 1 describes parameter settings of the MEC environment. The simulated MEC environment consists of three edge servers and one UAV as the end-device. Edge servers are deployed close to the UAV, the data transmission rate between UAV and edge server is 100 Mbps [30] and the bandwidth is relatively stable. The processing speed of the computing resources is randomly chosen between 100 and 1500 Megacycles. According to the EX-PRESS framework and actual data collection in UAV last-mile delivery scenarios, the input and output data size for each task is generated between 0.625 and 30 MB randomly [21], [31], [32]. In this paper, we only consider the energy consumption of the UAV.

Table 2 describes the parameter settings of PSO and GA strategy respectively [33], [34]. For each workflow application, we simulate 100 times to obtain the average result. In order to comprehensively evaluate the overall performance, we randomly generate many complex WDG of different sizes from 10 to 100 tasks with both linear and nonlinear workflow structures. The percentage of the local execution task (namely those cannot be offloaded to the edge) is set as 20%.

### TABLE 2
### PSO, GA STRATEFGY PARAMETER SETTING

| PSO | | GA | |
|---|---|---|---|
| Parameters | Setting | Parameters | Setting |
| Particles | 30 | Population Size | 50 |
| Iterations | 100 | Iterations | 100 |
| Factor C1, C2 | 2 | Cross Rate | 0.8 |
| Inertia Weight | 1 | Mutation Rate | 0.1 |
| Repeated | 10 | Repeated | 10 |

## 5.3 Performance Evaluation

Now we present the detailed simulation results. Deadline constraint is the most important QoS constraints in any real-world business systems. According to actual business requirements, there is usually a strict time constraint. Based on the results of our actual program [21], our workflow application's deadline constraint is set between 70% and 140% of the total task local execution time. Note that the deadline constraint considered in this paper is the soft deadline, which means that missing the deadline will not cause task failures but only decrease the service quality.

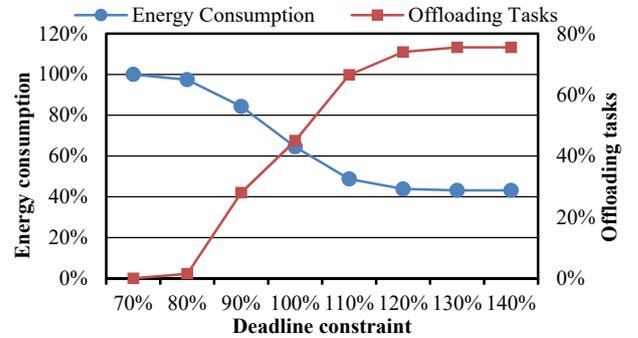

Fig. 12. Energy consumption and task's offloading percent with different deadline constraints

Fig. 12 shows the energy consumption results under different deadline constraints. Initially, all tasks are executed locally on the UAV. When the value of deadline constraint increases, the percentage of offloaded tasks also increases. In the meantime, the end-device's energy consumption is gradually decreasing. It can be seen that the end-device's energy consumption becomes stable when the deadline constraint reaches 130%. The results show that when the deadline constraint becomes more flexible, the room for the computation offloading strategy becomes larger and hence our Graph4Edge-Nonlinear strategy becomes more effective in reducing the end-device's energy consumption. However, when deadline constraint reaches a certain value, the effectiveness of reducing energy consumption levels off since the percentage of offloaded tasks also reaches its maximum 80% (as we have set 20% of local execution tasks). In order to compare the best performance of different computation offloading strategies, we focus on the deadline is 130% in the following experiments.

To comprehensively evaluate its performance, we compare Graph4Edge-Nonlinear with others in energy consumption, strategy running time and task response time.

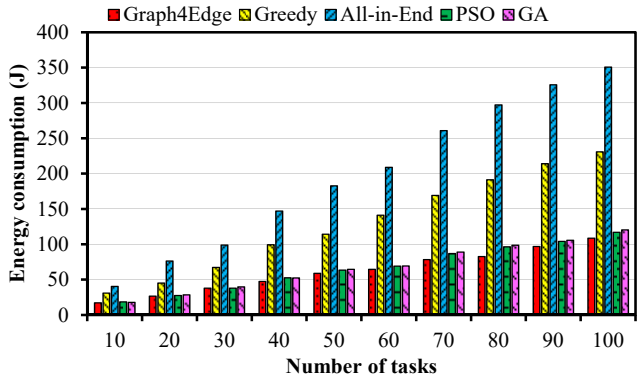

Fig. 13. Comparison of Graph4Edge-Nonlinear and other four methods for energy consumption

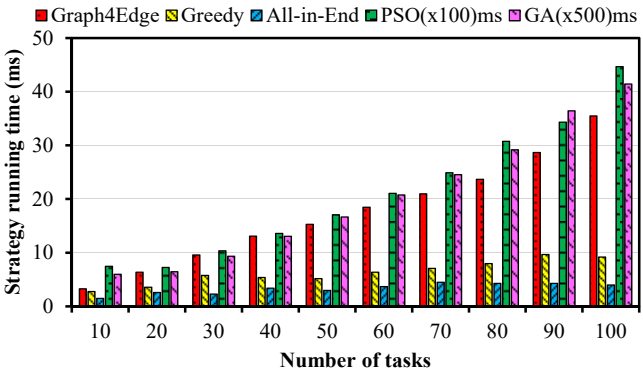

Fig. 14. Comparison of Graph4Edge-Nonlinear and other four methods for strategy running time

Fig. 13 shows that the end-device's energy consumption of the over different sizes of nonlinear workflows. The experimental result demonstrates that energy consumption with Graph4Edge-Nonlinear is always lower than the other four strategies. For example, when the task number is 50, the energy consumption with Graph4Edge-Nonlinear is 7.81% and 9.51% lower than PSO and GA respectively. The Greedy strategy and All-in-End strategy always have higher energy consumption than Graph4Edge-Nonlinear which is about 94% and 310% respectively. This is because they place too many tasks on the end device for execution.

Fig. 14 illustrates the results of strategy running time over different sizes of nonlinear workflows. Please be noted that the basic time units for Graph4Edge-Nonlinear, Greedy and the All-in-End strategies are milliseconds (ms), while the basic time units for PSO and GA are 100ms and 500ms respectively. Clearly, Graph4Edge-Nonlinear is much faster than search-based strategies PSO and GA. Specifically, it is running about 110 times and 540 times faster than PSO and GA respectively. In the FogWorkflowSim platform [22], All-in-End strategy running needs to count the number of tasks that make up the workflow, which usually takes a few milliseconds. Although the Greedy and All-in-End strategies have the smallest execution time, it always has the worst performance in reducing energy consumption. In particular, even if the number of tasks is as high as 100, the Graph4Edge strategy only needs an additional 30ms to get the optimal offloading decision. When it is necessary to perform offloading of delay-sensitive tasks, our strategy can guarantee real-time performance.

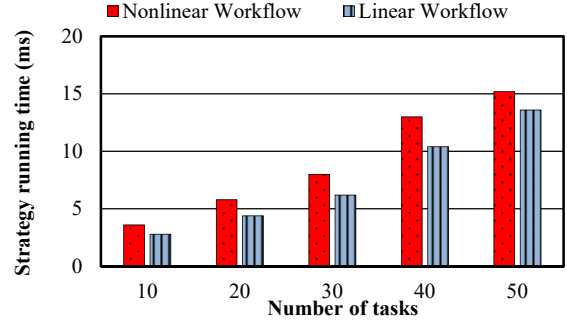

Fig. 15. Comparison of nonlinear workflows and linear workflows for strategy running time

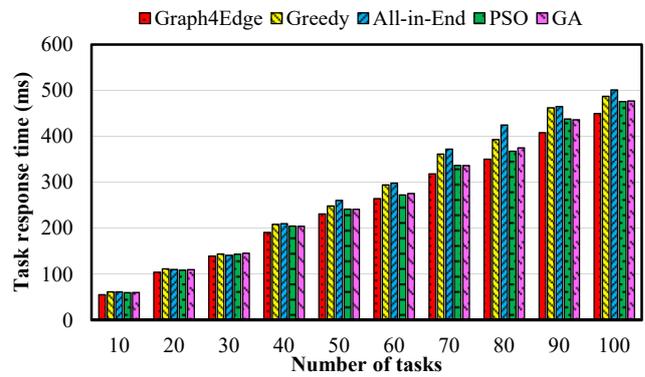

Fig. 16. Comparison of Graph4Edge-Nonlinear and other four methods for task response time

It is also important to investigate the impact of workflow structures on the strategy running time. Fig. 15 compares the strategy running time of Graph4Edge-Nonlinear and Graph4Edge-Linear on nonlinear workflows and linear workflows respectively with 10 to 50 tasks. The result shows that with the same workflow sizes, the strategy running time of Graph4Edge-Nonlinear is about 20% higher than Graph4Edge-Linear, which is not a significant increase considering the much more complex structures of the nonlinear workflows. Meanwhile, even with 50 tasks, the strategy running time of Graph4Edge-Nonlinear is only increased by 2ms.

Fig. 16 illustrates the results of task response time over different sizes of tasks in nonlinear workflows. The workflow tasks are executed according to the offloading decision plan. The task response time is set as the sum of task execution time and data transmission time. Obviously, our proposed Graph4Edge-Nonlinear strategy requires minimal task response time. In summary, given the experimental results above, we can conclude that our proposed strategy has the best overall performance as it can find the optimal computation offloading decision plan with the lowest energy consumption under the given deadline. Most importantly, for delay-sensitive applications, our proposed strategy can meet their real-time requirements given its fast running time.

# 6 RELATED WORK

In a MEC environment, the main objective of the computation offloading strategy is to find the best offloading decision according to the characteristics of tasks, computing resources and network conditions [35]. In addition, these methods can improve the utilization rate of computing resources in MEC [36] and minimize the end-device's energy consumption with the QoS constraints of users and reduce the cost of the service providers [37]. Computation offloading is used for solving the problems caused by insufficient computing power and insufficient battery capacity of the end-device. With the gradual promotion of MEC platform, computation offloading has become an important research topic [25]. There are many preliminary research works focusing on the problem of computation offloading in the MEC environment.

Currently, a range of research works focuses on the problem of computation offloading to reduce end-device's energy consumption and task response time. In the aspect of the energy consumption optimization, Zhang et al. [38] focused on the multi-access characteristics of the 5G and an minimize the energy consumption offloading strategy in 5G MEC networks is proposed. Chen et al. [39] consider the factor of task allocation and CPU-cycle frequency. Then, an energy efficiency strategy named TOFFEE is proposed to decrease end-device's energy consumption. For the target of time optimization, Xing et al. [40] propose a MEC system that can reduce the task's computation latency significantly. Currently, many research works are based on popular optimization algorithms such as PSO and GA [33], [34]. However, most studies only focused on the independent tasks, and ignored the computational overhead needed to make decisions. Most of the tasks in real-world applications are highly correlated. Each task depends on the execution result of the previous task and provides the necessary data flow for the successor task [31], [41].

Graph as a kind of classical data structure. Many research works about graph-based algorithms in Cloud and Edge Computing. In the cloud computing environment, many studies adopted graph-based algorithms to solve task management problems for different workflow structures. Yuan et al. [42] aimed at the problem of data set storage in data-intensive scientific workflow execution and proposed the CTT-SP algorithm to trade-off computation and storage cost in the cloud environment. Zhang et al. [43] proposed a highly efficient algorithm named PCE, which can calculate the minimum cost strategy in multiple clouds efficiently. In recent years, there are also many studies in the MEC environment. Elgamal et al. [44] focused on the time and memory of the pricing model in serverless computing and proposed an algorithm to optimize the price of AWS workflow applications while meeting deadline constraints. Khare et al. [45] focused on the data placement problem of operators in streaming applications and proposed an algorithm to convert streaming DAG into a set of approximate linear chains and perform data placement and time prediction. Most research focuses on resource management issues in cloud or edge conditions. However, these works have not paid attention to the problems of end-device's limited computing power and battery capacity in the MEC environment. In [25], the authors proposed a graph-based strategy that can provide a solution for the generation of the optimal energy consumption offloading strategy for linear workflow, but this strategy cannot handle the nonlinear workflow of complex applications.

Therefore, this paper deals with various problems of computation offloading in the MEC environment using graph-based technology. We focus on workflow applications with complex structures and propose the Graph4Edge-Nonlinear strategy as an effective solution.

# 7 CONCLUSION AND FUTURE WORK

Computation offloading is a key technology to optimize the QoS of MEC based applications. However, most existing strategies did not pay attention to the dependency between computing tasks or only focus on simple dependency such as sequential relationship. Meanwhile, many current strategies are based on search algorithms which could have significant computation overhead. This is unacceptable for delay-sensitive applications. For solving these problems, a novel graph-based computation offloading strategy with the goal to minimize the end-device's energy consumption under the given deadline constraint is proposed in this paper. Motivated by a MEC-based UAV delivery system, we first built the nonlinear workflow model for complex applications. Then, using the graph-based partition technique, we proposed the Graph4Edge-Nonlinear strategy to search for the best computation offloading decision with the lowest energy consumption under the deadline constraint. Finally, both a real-world case study and comprehensive simulation experiments implemented on the FogWorkflowSim platform with different workflow sizes and structures are conducted to evaluate the effectiveness of our proposed strategy. The experimental results have shown that Graph4Edge-Nonlinear can achieve overall better performance than other representative computation offloading strategies.

This paper mainly focused on computation offloading for workflow applications and the target of reducing end-device's energy consumption in the MEC environment. In the future, we will investigate the problem of computation offloading together with workflow scheduling at the edge servers to produce a holistic solution that can improve the QoS for the whole MEC-based system. Besides, the impact of the dynamics of the network and the mobility of the end device will be further investigated in our future work.

## ACKNOWLEDGMENT

This work was supported by the National Natural Science Foundation of China (No.61972001, No.62076002), the National Natural Science Foundation of Anhui Province (No. 2008085MF194) and in part by the Humanities and Social Sciences of MOE Project No. 16YJCZH048.

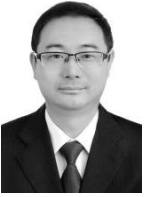

**Xuejun Li** (Member, IEEE) received the Ph.D. degree in computer application technology from the School of Computer Science and Technology, Anhui University, Hefei, Anhui, China, in 2008. He is currently a Full Professor with the School of Computer Science and Technology, Anhui University, Hefei, Anhui, China. His major research interests include mobile edge computing, workflow systems, cloud computing, and intelligent software.

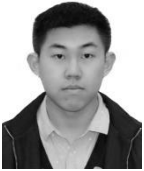

**Tianxiang Chen** received the bachelor's degree in Internet of Things technology from the School of Computer Science and Technology, Fuyang Normal University, Fuyang, Anhui, China, in 2018. He is currently pursuing the master's degree with the School of Computer Science and Technology, Anhui University, Hefei, Anhui, China. His current research interests include mobile edge computing, workflow system, cloud computing, resource management.

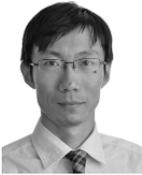

**Dong Yuan** (Member, IEEE) received the BEng and MEng degrees from Shandong University, Jinan, China, in 2005 and 2008, respectively, and the PhD degree from Swinburne University of Technology, Melbourne, Australia, in 2012, all in computer science. He is a senior lecturer with the School of Electrical and Information Engineering, the University of Sydney, Sydney, Australia. His research interests include cloud computing, parallel and distributed systems, scheduling and resource management, deep learning, data management and Internet of Things.

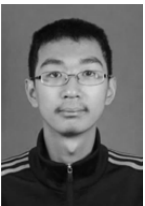

**Jia Xu** (Student Member, IEEE) received the bachelor's and master's degree in computer science and technology from the School of Computer Science and Technology, Anhui University, Hefei, Anhui, China, in 2010-2017, respectively. He is currently pursuing the Ph.D. degree with the School of Computer Science and Technology, Anhui University, Hefei, Anhui, China. He was a Software Engineer focusing on industrial projects and solutions in iFLYTEK Co., Ltm from 2017-2018. His current research interests include mobile edge computing, workflow system, cloud computing, resource management.

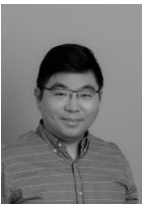

**Xiao Liu** (Senior Member, IEEE) received the bachelor's and master's degrees in information management and information system from the School of Management, Hefei University of Technology, Hefei, China, in 2004 and 2007, respectively, and the Ph.D. degree in computer science and software engineering from the Faculty of Information and Communication Technologies, Swinburne University of Technology, Melbourne, Australia, in 2011. He was teaching at the Software Engineering Institute, East China Normal University, Shanghai, China. He is currently a Senior Lecturer with the School of Information Technology, Deakin University, Melbourne. His current research interests include software engineering, distributed computing, and data mining, with special interests in workflow systems, cloud/fog computing, and social networks.